\documentclass[%
 reprint,
 amsmath,amssymb,
aps,
prl,
]{revtex4-1}

\usepackage{graphicx}% Include figure files
\usepackage{dcolumn}% Align table columns on decimal point
\usepackage{bm}% bold math
\newcommand{\formref}[1]{(\ref{#1})}
\newcommand{\brac}[1]{\left(#1\right)}

\newcommand{\bracm}[1]{\left\langle #1\right\rangle}
\newcommand{\abs}[1]{\left|#1\right|}

\newcommand{\cet}[1]{\left|#1\right\rangle}
\newcommand{\bra}[1]{\left\langle#1\right|}

\begin{document}

\preprint{APS/123-QED}

\title
%{Quantum Langevin noise in a hot resonantly driven atomic medium}
{The fluctuation-dissipation relation in a resonantly driven quantum medium}

% Force line breaks with \\
%\thanks{A footnote to the article title}%

\author{Maria Erukhimova}
\email{eruhmary@appl.sci-nnov.ru}
 %\altaffiliation[Also at ]{Physics Department, XYZ University.}%Lines break automatically or can be forced with \\
\author{Mikhail Tokman}%
 %\email{Second.Author@institution.edu}
\affiliation{Institute of Applied Physics RAS, Uljanova str, 46,
Nizhny Novgorod, Russia}

\date{\today}% It is always \today, today,
             %  but any date may be explicitly specified

\begin{abstract}

We calculate the radiation noise level associated with the spontaneous emission of a coherently driven medium. The significant field-induced modification of relation between the noise power and damping constant in a thermal reservoir is obtained. The nonlinear noise exchange between different atomic frequencies leads to violation of standard relations dictated by the fluctuation-dissipation theorem.

\begin{description}
\item[PACS numbers] 42.50.Lc, 42.50.Ct, 42.50.Gy

%May be entered using the 
%\verb+\pacs{#1}+ command.
\end{description}
\end{abstract}

\pacs{Valid PACS appear here}% PACS, the Physics and Astronomy
                             % Classification Scheme.
%\keywords{Suggested keywords}%Use showkeys class option if keyword
                              %display desired
\maketitle

%\section{\label{Introduction}Introduction}

{\it Introduction -- } 
%The most of known methods of manipulation of quantum statistics of light are based on nonlinear optical processes. Preparation of nonclassical states of light (squeezed, entangled  and so on) attracts the most attention \cite{Scully}. Such states are distinguished by extremely low dispersion of some physical quantities. The additional noise having a thermal nature associated with the processes of spontaneous emission in the environment is a major factor that destroys the ideal quantum state of light.It seems to be very important to develop a correct method of estimation of thermal noise source characteristics in complicated nonlinear systems going beyond the simple estimates based on Fluctuation-Dissipation Theorem (FDT) in its standard variant that is valid for linear dissipative media. In present letter we theoretically predict the effect of drastic field induced modification of thermal noise source for quantum radiation.
Investigation of fluctuations in open systems (both quantum and classical) is important in many different areas of physics being the subject of incessant general physical interest \cite{Landau_Stat_Phys, Pokrovsky, Jarzynski, Marconi, Maghrebi}. One of the major results in this field is the Fluctuation-Dissipation Theorem (FDT) \cite{Callen}. The use of the universal relation between relaxation and fluctuation properties of the medium imposed by FDT turned out to be highly fruitful in the theory of radiation-matter interaction \cite{Rytov}. But in its "standard" variant the FDT is not applicable for nonlinear processes that are of great interest. The theory developed in \cite{Fain} for calculation of polarisation fluctuations in the presence of coherent drive at the combinational frequency seems to be not applicable for the case of resonant interaction with the medium since it is based on the methods of perturbation theory.
%, and besides it does not take into account the finite relaxation times in the medium.
Meanwhile one of the most interesting applications of the theory of fluctuations is connected specifically with the resonant control of quantum medium. A number of attractive effects of quantum optics, based on preparation and manipulation of nonclassical states of light with the extremely low dispersion of some physical quantities (squeezed, entangled  and so on) \cite{Scully} are related to these processes. 
%It seems that fluctuation theory in such systems should combine formulation of universal relations and methods (similar to FDT) with peculiarities of particular systems taking into account.       
%Preserving commutation relations for the field operators in media with significant dissipation (that is especially important for resonance processes) requires taking into account the noise Langevin sources. 
The detailed analysis of Langevin sources influence on the preparation of nonclassical states of light in different nonlinear resonance regimes have been carried out in a wide range of papers \cite{Shwartz, Lukin_Matsko, Balic, Kolchin07, Kolchin06, Glorieux, Lauk}. But in most of papers the consideration was restricted to the case of zero temperature. So only the principal possibility of suppression of intrinsic quantum fluctuations in dissipative media was investigated. The additional noise having a thermal nature associated with the processes of spontaneous emission in the environment is a major factor that destroys the ideal quantum state of light. The possibility of suppression (squeezing) of "thermal" fluctuations due to the same mechanisms of nonlinear interactions was considered in the paper \cite{Galve}, but the authors used a simplified method estimating the noise source as for non-interacting oscillators. 

In this Letter we develop the general method for calculation of correlation characteristics of the noise Langevin sources for a quantum field in a coherently driven resonant dissipative atomic medium being in contact with a thermal reservoir. The developed theory combines formulation of universal relations (similar to FDT) with peculiarities of specific systems. We apply this method to consider the \(\Lambda-\)scheme of Electromagnetically Induced Transparency (EIT) which is the basis for many mechanisms of nonclassical light manipulation (see, for example \cite{Vdovin, Lukin_Matsko, Balic, Kolchin07, Kolchin06,  McCormick, Glorieux}). We theoretically predict the effect of drastic field induced modification of thermal noise source for quantum radiation.
%We theoretically predict the effect of drastic field induced modification of thermal noise source for quantum radiation.

{\it Formalism --} 
We consider the interaction of a signal photon field with an atomic medium. 
The atoms have energy levels \(W_m\) (corresponding energy states \(\cet{m}\)). 
They can be driven by any classical coherent field. 
%It is considered as given and undepleted. 
The atomic system is open being under the influence of the dissipative reservoir. The atoms-quantum field system is described by the set of coupled Heisenberg-Langevin equations. Using continuous medium approximation we describe the ensemble of atoms  with collective coordinate-dependent density operators \cite{Kolchin07, Glorieux, Tokman_Yao_Belyanin} defined as 
\begin{equation}
\hat{\rho}_{mn}\brac{\mathbf{r},t}=\frac{1}{\Delta V_r}\sum_j\hat{\rho}_{j;mn}(t),
\end{equation}
where index \(j\) numerates the atoms within the small volume \(\Delta V_r\) in the vicinity of point with radius-vector \(\mathbf{r}\), 
\(\hat{\rho}_{j;mn}=\cet{n}_j\bra{m}_j\) 
is the Heisenberg projection operator 
acting on variables of atom with index \(j\).
Within the Heisenberg-Langevin approach the operator \(\hat{\rho}_{mn}\) obeys the equation \cite{Tokman_Yao_Belyanin}: 
\begin{equation}
\dot{\hat{\rho}}_{mn}
=-\frac{i}{\hslash}\brac{\hat{h}_{mp}\hat{\rho}_{pn}-\hat{\rho}_{mp}\hat{h}_{pn}}+\hat{R}_{mn}+\hat{F}_{mn},
\label{rho}
\end{equation}  
where \(\hat{h}_{mn}=W_m\delta_{mn}-\mathbf{d}_{mn}\hat{\mathbf{E}}\brac{\mathbf{r},t}\) is the matrix element of the Hamiltonian operator taking into account interaction of atoms with the electric field \(\hat{\mathbf{E}}\brac{\mathbf{r},t}\) in the dipole approximation, \(\hat{R}_{mn}\) is the relaxation operator, and \(\hat{F}_{mn}\) is the Langevin noise operator satisfying \(\bracm{\hat{F}_{mn}}=0\), where averaging is taken over the reservoir variables. 

The electric field operator \(\hat{\mathbf{E}}\brac{\mathbf{r},t}\) obeys the equation (see for example \cite{Fain, Tokman_Yao_Belyanin, Andreev}):
\begin{equation}
\frac{\partial^2}{\partial t^2}\hat{\mathbf{E}}+c^2\nabla\times\nabla\times\hat{\mathbf{E}}=
-4\pi\frac{\partial^2}{\partial t^2}\hat{\mathbf{P}},
\end{equation}
where the operator of electric polarization is expressed via the density operators as \(\hat{\mathbf{P}}\brac{\mathbf{r},t}=\sum_{m,n}\mathbf{d}_{nm}\hat{\rho}_{mn}\), which in turn are the solution of the Eq.~\formref{rho}.

For a wide range of regimes of interaction between quantum radiation and dense atomic medium the linear approximation is applicable, whereas the nonlinear response, as well as Langevin effects can be taken into account as small additional terms to the linear relation \cite{Tokman_Yao_Belyanin, Vdovin}:
\begin{eqnarray}
%\hat{\mathbf{P}}\brac{\mathbf{r},t}=\int_0^{\infty}\chi^H(\tau)\hat{\mathbf{E}}\brac{\mathbf{r},t-\tau}d\tau+\delta\hat{\mathbf{P}}_{NL}+\delta\hat{\mathbf{P}}_{diss}+\delta\hat{\mathbf{P}}_{L}.
\hat{\mathbf{P}}\brac{\mathbf{r},t}=\int_0^{\infty}\brac{\chi^H(\tau)+\chi^{aH}(\tau)}\hat{\mathbf{E}}\brac{\mathbf{r},t-\tau}d\tau\nonumber\\+\delta\hat{\mathbf{P}}_{NL}
+\delta\hat{\mathbf{P}}_{L}.
\label{P}
\end{eqnarray}
The functions \(\chi^H(\omega)=\int_0^\infty{\chi^H(\tau)}e^{i\omega \tau}d\tau\), \(\chi^{aH}(\omega)=\int_0^\infty{\chi^{aH}(\tau)}e^{i\omega \tau}d\tau\) 
presents the Hermitian and anti-Hermitian components of the susceptibility of a medium. 

The Fluctuation-Dissipation Theorem (FDT) in its standard form \cite{Fain, Rytov} defines the relation between the spectral density of fluctuations of polarisation in the medium with anti-hermitian component of the medium's susceptibility:
\begin{eqnarray}
\frac{1}{2}\bracm{\delta\hat{\mathbf{P}}_{L}^\dag\brac{\mathbf{r},\omega}\delta\hat{\mathbf{P}}_{L}\brac{\mathbf{r'},\omega'}+\delta\hat{\mathbf{P}}_{L}\brac{\mathbf{r'},\omega'}\delta\hat{\mathbf{P}}_{L}^\dag\brac{\mathbf{r},\omega}}
=\nonumber\\-i\frac{\hbar}{\pi}\chi^{aH}(\omega)\brac{n_T\brac{\omega}+\frac{1}{2}}\delta\brac{\omega-\omega'}\delta\brac{\mathbf{r}-\mathbf{r'}},\nonumber\\
\label{fdt1}
\end{eqnarray}
where \(n_T\brac{\omega}=\brac{e^{\hbar\omega/T}-1}^{-1}\) is the averaged number of "thermal" photons at frequency \(\omega\). This relation is true for the unperturbed stationary medium which is in thermal equilibrium with the reservoir at temperature \(T\) with the additional assumption of delta-correlation in space.

The question arises: can the FDT relation Eqs.~\formref{fdt1} be used for the coherently driven medium? 
We show that it is not sufficient just to take into account the drive-induced modification of dissipation properties at the operating frequency (drive intensity dependence \(\chi^{aH}(I_d)\)) and drive-induced pumping (effective change of the medium temperature). The modification of Eq.~\formref{fdt1} may be much more significant even for the case of relatively weak driving, much less than the saturation level. And the EIT system represents a shining example of that.

To obtain the correlation functions for the noise polarization \(\delta\hat{\mathbf{P}}_{L}\), we start with the analysis of correlation properties of Langevin operators in atomic equations Eq.~\formref{rho} \(\hat{F}_{mn}\) ("noise forces"). Then calculating the noise response in a medium we finally  get the noise polarisation that is the Langevin source for the quantum field.     
Note that we ignore the influence of any other, independent of atomic system, reservoir on the quantum field, supposing that in any case it can be considered additively and it does not depend on the drive action.

{\it Correlations of atomic Langevin operators -- }
The important point is that the properties of Langevin noise operators \(\hat{F}_{mn}\) in Eq.~\formref{rho} are connected with the properties of relaxation operators \(\hat{R}_{mn}\) \cite{Weiss}. The standard model for the relaxation operators  corresponds to the so-called Bloch-Redfield equations \cite{Weiss}, where 
\begin{equation}
\hat{R}_{mn}=\sum_{pq}r_{mnpq}\hat{\rho}_{pq}.
\label{r_mnpq}
\end{equation} 
In a simplest form the nonzero coefficients \(r_{mnpq}\) set the rates of transverse and longitudinal relaxation:
\begin{eqnarray}
\hat{R}_{mn}=-\gamma_{mn}\hat{\rho}_{mn}, m\neq n, \nonumber \\
\hat{R}_{mm}=\sum_n{w_{mn}\hat{\rho}_{nn}}, w_{mm}=-\sum_{m\neq n}w_{nm}.
\label{gamma_mn}
\end{eqnarray} 
The model of constant relaxation rates is in definite sense equivalent to the condition of \(\delta\)-correlated in time noise source. It is the result of the Markov approximation \cite{Weiss}. So that we can use the expression:
\begin{equation}
\bracm{\hat{F}_{mn}(\mathbf{r},t)\hat{F}_{pq}(\mathbf{r'},t')}=2D_{mnpq}\brac{\mathbf{r},\mathbf{r'},t}\delta\brac{t-t'}.
\label{dt}
\end{equation} 
This approximation allows one to use so-called generalized Einstein relations \cite{Scully,Cohen-Tannoudji} for calculating the diffusion coefficients \(D_{mnpq}\brac{\mathbf{r},\mathbf{r'},t}\):
\begin{eqnarray}
2D_{mnpg}\brac{\mathbf{r},\mathbf{r'}}=\frac{d}{dt}\bracm{\hat{\rho}_{mn}(\mathbf{r},t)\hat{\rho}_{pq}(\mathbf{r'},t)}-\nonumber\\
-\bracm{\brac{\frac{d}{dt}\hat{\rho}_{mn}(\mathbf{r},t)-\hat{F}_{mn}(\mathbf{r},t)}\hat{\rho}_{pq}(\mathbf{r'},t)}-\nonumber\\
-\bracm{\hat{\rho}_{mn}(\mathbf{r},t)\brac{\frac{d}{dt}\hat{\rho}_{pq}(\mathbf{r'},t)-\hat{F}_{pq}(\mathbf{r'},t)}}.
\label{GER}
\end{eqnarray}
Next, we assume that the action of reservoir on different atoms is independent, so that fluctuations of density matrix operators for different atoms are not correlated \(\bracm{\brac{\hat{\rho}_{mn;j}-\bracm{\hat{\rho}_{mn;j}}}\brac{\hat{\rho}_{pq;i}-\bracm{\hat{\rho}_{pq;i}}}}\propto\delta_{ij}\).
Taking into account also the strict equality, that should be fulfilled for each atom by definition of density matrix operators: \(\hat{\rho}_{mn;j}\hat{\rho}_{pq;j}=\hat{\rho}_{pn;j}\delta_{mq}\), we get for the averaged product of space-dependent density matrix operators the following relation:
\begin{eqnarray}
\bracm{\hat{\rho}_{mn}(\mathbf{r})\hat{\rho}_{pq}(\mathbf{r'})}=\nonumber\\
\bracm{\hat{\rho}_{mn}(\mathbf{r})}\bracm{\hat{\rho}_{pq}(\mathbf{r'})}+\delta_{mq}\bracm{\hat{\rho}_{pn}(\mathbf{r})}\delta(\mathbf{r}-\mathbf{r'}).
\label{product_rho}
\end{eqnarray}
The expression for the diffusion coefficients Eq.~\formref{GER} with regard to Eqs.~\formref{product_rho},\formref{rho},\formref{r_mnpq} finally takes \(\delta\)-correlated in space form:  
\begin{equation}
D_{mnpq}(\mathbf{r},\mathbf{r'},t)=D_{mnpq}(\mathbf{r},t)\delta(\mathbf{r}-\mathbf{r}'),
\label{dr}
\end{equation}
where 
\begin{eqnarray}
2D_{mnpq}(\mathbf{r},t)=\nonumber\\
\delta_{mq}\bracm{\hat{R}_{pn}}-\sum_{l}r_{mnql}\bracm{\hat{\rho}_{pl}}-\sum_{k}r_{pqkm}\bracm{\hat{\rho}_{kn}}.
\end{eqnarray}
In particular, the correlation functions of the Langevin sources for the "off-diagonal" operators (\(m\neq n\),\(p\neq q\)) taking into account Eq.~\formref{gamma_mn} are given by a simple expression:
\begin{equation}
2D_{mnpq}(\mathbf{r},t)=\delta_{mq}\brac{\brac{\gamma_{mn}+\gamma_{pq}}\bracm{\hat{\rho}_{pn}}+\bracm{\hat{R}_{pn}}}.
\label{D_mnpq}
\end{equation}

With the final aim to calculate the noise polarization \(\delta\hat{\mathbf{P}}_L\) as the noise source for the quantum radiation Eq.~\formref{P} we use the approximation when the atomic response at the Langevin force can be found neglecting quantum field action.
At the same time the atomic system is assumed to be driven by a classical external field \(\mathbf{E_d}=\mathbf{e_d}E_dexp(-i\omega_d t+i\mathbf{k_d}\mathbf{r})+c.c.\), that excites off-diagonal density matrix operators 
and modify diagonal ones. So that the properties of atomic Langevin operators, defined by expression Eq.~\formref{D_mnpq}, depend on the drive wave.
(This dependence can be more complicated if we take into account the field induced modification of relaxation parameters \cite{Kocharovskaya, Radeonychev}. However we neglect this effect here.) 
Thus, the autocorrelation function for Langevin operator at some transition \(m-n\) is modified due to field-induced redistribution of populations:
\begin{eqnarray}
2D_{mnnm}(\mathbf{r},t)=2\gamma_{mn}\bracm{\hat{\rho}_{nn}}+\bracm{\hat{R}_{nn}}.
\label{D_mnnm}
\end{eqnarray}
Besides, the excited coherence at some transition \(a-b\) generates the non-zero correlations of Langevin sources at adjacent atomic transitions:
\begin{eqnarray}
2D_{mabm}(\mathbf{r},t)=\brac{\gamma_{am}+\gamma_{bm}-\gamma_{ab}}\bracm{\hat{\rho}_{ba}}.
\label{D_mabm}
\end{eqnarray} 

It is useful to derive the expression for the spectral components of the Langevin operators \(\hat{F}_{mn}\brac{\mathbf{r},t}=\int_{-\infty}^{+\infty}{\hat{F}_{mn}\brac{\mathbf{r},\omega}e^{-i\omega t}d\omega}\).
Under the approximation of constant populations and amplitude of drive-induced coherence in resonant approximation \(\bracm{\hat{\rho}_{nn}}=const\), \(\bracm{\hat{\rho}_{ba}}=\left.{\sigma_{ba}e^{\mp i\omega_d t}}\right|_{b\gtrless a}\), \(\sigma_{ba}=const\), we get from Eq.~\formref{D_mnnm} and Eq.~\formref{D_mabm}:
\begin{eqnarray}
\bracm{\hat{F}_{mn}(\mathbf{r},\omega)\hat{F}_{nm}(\mathbf{r'},\omega')}=\nonumber\\
\frac{1}{2\pi}\brac{2\gamma_{mn}\bracm{\hat{\rho}_{nn}}+\bracm{\hat{R}_{nn}}}\delta(\omega+\omega')\delta(\mathbf{r}-\mathbf{r}')\nonumber\\
\label{mnnm_omega}\\
\bracm{\hat{F}_{ma}(\mathbf{r},\omega)\hat{F}_{bm}(\mathbf{r'},\omega')}=
\frac{1}{2\pi}\brac{\gamma_{am}+\gamma_{bm}-\gamma_{ab}}\times\nonumber\\
\left.{\sigma_{ba}\delta(\omega+\omega'\mp \omega_d)}\right|_{b\gtrless a}\delta(\mathbf{r}-\mathbf{r}').\nonumber\\
\label{mabm_omega}
\end{eqnarray}

{\it Noise polarization in the EIT scheme -- }
\begin{figure}
\includegraphics[width=0.6\columnwidth, keepaspectratio=true]{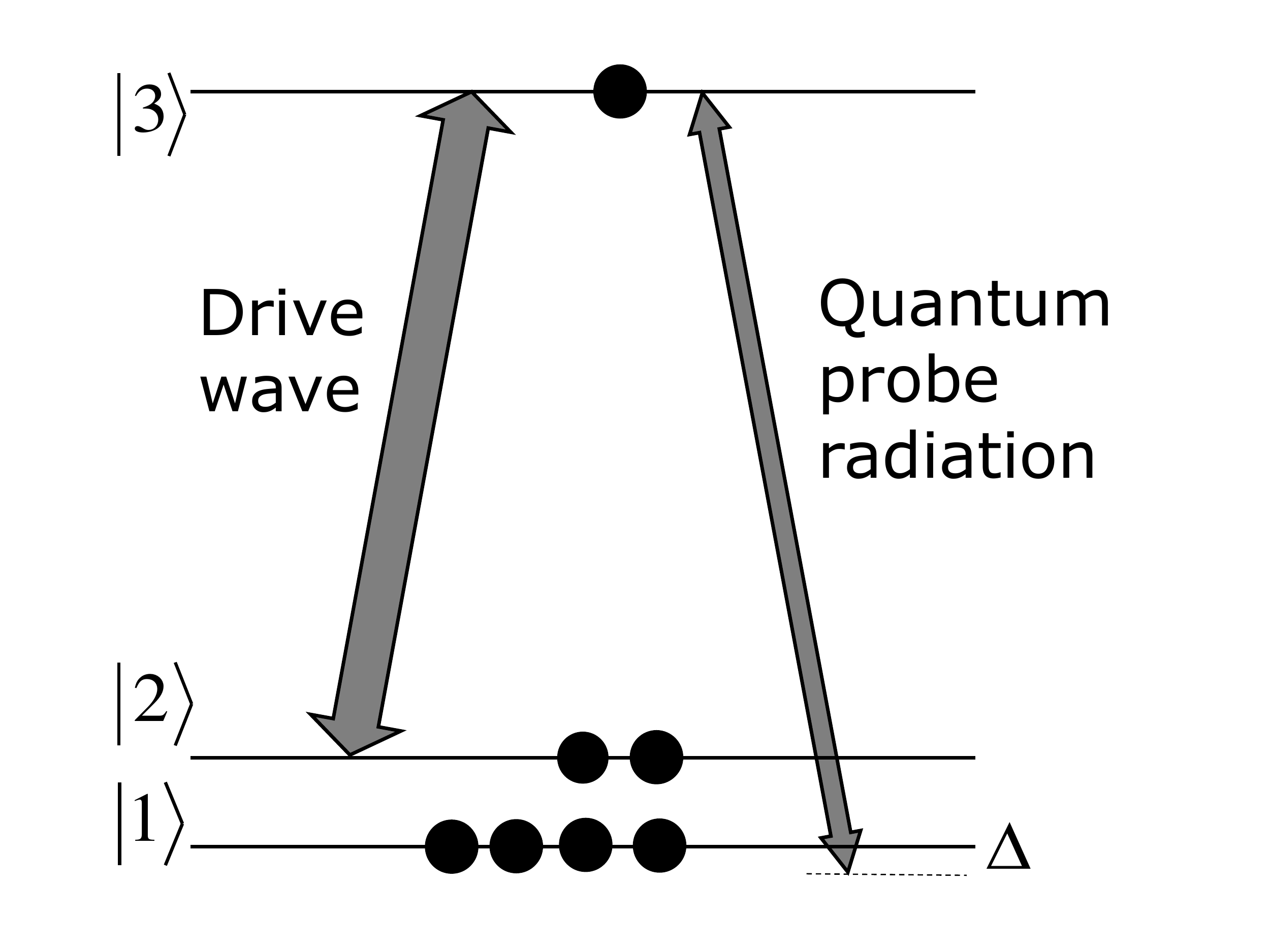}
\caption{\label{scheme} \(\Lambda\)-scheme of EIT.}
\end{figure}
Apply the obtained relations Eq.~\formref{mnnm_omega},\formref{mabm_omega} to the calculation of the noise source for quantum radiation, interacting with three-level atoms under EIT conditions (see Fig.~\ref{scheme}).
 We should solve equations for density matrix operators Eq.~\formref{rho} in the presence of drive and probe fields  \(\hat{\mathbf{E}}=\brac{\mathbf{e_d}E_d e^{i\mathbf{k_d}\mathbf{r}-i\omega_d t}+c.c.}+\brac{
\mathbf{e_p}\hat{E}_p\brac{\mathbf{r},t}e^{i\mathbf{k_p}\mathbf{r}-i\omega_p t}+H.c.}\) and Langevin forces \(\hat{F}_{mn}\). The frequency of drive wave is close (for simplicity equal) to the frequency of transition \(\cet{2}-\cet{3}\): \(\omega_d=\omega_{32}\). 
The quantum radiation spectrum is assumed to be narrow in scale of atomic frequencies with the center at \(\omega_p=\omega_{31}+\Delta_p\). We apply the RWA approximation and operate with slow variables:  \(\hat{\rho}_{32,31,21}=\hat{\sigma}_{32,31,21}e^{-i\omega_{d,p,l}t+i\mathbf{k}_{d,p,l}\mathbf{r}}\), \(\hat{F}_{32,31,21}=\hat{f}_{32,31,21}e^{-i\omega_{d,p,l}t+i\mathbf{k}_{d,p,l}\mathbf{r}}\), where \(\omega_l=\omega_p-\omega_d\), \(\mathbf{k}_l=\mathbf{k}_p-\mathbf{k}_d\).
The spectral component of polarization at probe wave frequency band is expressed via spectral component of coherence at transition \(\cet{3}-\cet{1}\): \(\hat{\mathbf{P}}_{\omega}=\mathbf{\hat{d}}_{13}\hat{\sigma}_{31,\nu=\omega-\omega_{p}}\). 
The atomic response to the action of quantum radiation and noise force in linear approximation can be considered additively: \(\hat{\sigma}_{31,\nu}=\hat{\sigma}_{31,\nu}^{qr}+\hat{\sigma}_{31,\nu}^{f}\). The corresponding solution depends on averaged values \(\bracm{\hat{\rho}_{ii}}=\rho_{ii}\) and \(\bracm{\hat{\sigma}_{32}}=\sigma_{32}\), that can be found as stationary solutions of corresponding equations, in particular \(\sigma_{32}=i\Omega_d n_{23}/\gamma_{32}\), where \(n_{23}=\rho_{22}-\rho_{33}\), \(\Omega_d=\mathbf{d_{32}}\mathbf{e_d}E_d/\hbar\). As a result we get the following solution of Eq.~\formref{rho}:
\begin{eqnarray}
\hat{\sigma}_{31,\nu}^{qr}=i\hat{\Omega}_p\frac{n_{13}\brac{\gamma_{21}-i(\Delta_p+\nu)}-n_{23}\abs{\Omega_d}^2/\gamma_{32}}{\abs{\Omega_d}^2+\brac{\gamma_{21}-i(\Delta_p+\nu)}\brac{\gamma_{31}-i(\Delta_p+\nu)}}\nonumber\\\label{sigma^qr}\\
\hat{\sigma}_{31,\nu}^{f}=\frac{\brac{\gamma_{21}-i\brac{\Delta_p+\nu}}\hat{f}_{31,\nu}+i\Omega_d \hat{f}_{21,\nu}}{\abs{\Omega_d}^2+\brac{\gamma_{21}-i(\Delta_p+\nu)}\brac{\gamma_{31}-i(\Delta_p+\nu)}}.\nonumber\\\label{sigma^f}
\end{eqnarray}
From the first of this expressions Eq.~\formref{sigma^qr} one gets the dielectric susceptibility at frequency \(\omega\approx\omega_p\) of coherently driven medium with its anti-Hermitian part:
\begin{widetext}
\begin{equation}
\chi^{aH}(\omega,\Omega_d)=\frac{i\abs{d_{31}}^2}{\hbar}\frac{n_{13}\brac{\gamma_{21}\abs{\Omega_d}^2+\gamma_{31}\brac{\gamma_{21}^2+\Delta^2}}-\frac{\abs{\Omega_d}^2n_{23}}{\gamma_{32}}\brac{\abs{\Omega_d}^2+\gamma_{32}\gamma_{31}-\Delta^2}}{\brac{\abs{\Omega_d}^2+\gamma_{21}\gamma_{31}-\Delta^2}^2+\Delta^2\brac{\gamma_{21}+\gamma_{31}}^2},
\label{chi_Omega}
\end{equation} 
\end{widetext}
here \(\Delta=\omega-\omega_{31}\). From the second equation Eq.~\formref{sigma^f} with the use of Eqs.~\formref{mnnm_omega},\formref{mabm_omega} we calculate the magnitude of the correlation function for noise polarisation at probe frequency. The expression Eq.~\formref{sigma^f} shows that under the condition of clear-cut EIT, that is
\begin{equation}
\gamma_{31}^2\gg\abs{\Omega_d}^2\gg\gamma_{21}\gamma_{31},
\label{EIT_condition}
\end{equation}
the intensity of noise source for the radiation at frequency \(\omega_p\) will depend not only and not so much on the "noise force" at the resonant transition \(\cet{3}-\cet{1}\) but on the parametrically transferred "noise force" acting at the adjacent low frequency transition \(\cet{2}-\cet{1}\) and correlations between these forces.
Taking into account \(\bracm{R_{22}}=-\bracm{R_{33}}=2 Im(\Omega_d^*\sigma_{32})\), \(\bracm{R_{11}}=0\) we finally get:
\begin{eqnarray}
\bracm{\delta\hat{\mathbf{P}}_{L}\brac{\mathbf{r'},\omega'}\delta\hat{\mathbf{P}}_{L}^\dag\brac{\mathbf{r},\omega}-\delta\hat{\mathbf{P}}_{L}^\dag\brac{\mathbf{r},\omega}\delta\hat{\mathbf{P}}_{L}\brac{\mathbf{r'},\omega'}}=\nonumber\\-i\frac{\hbar}{\pi}\chi^{aH}\brac{\omega,\Omega_d}\delta\brac{\omega-\omega'}\delta\brac{\mathbf{r}-\mathbf{r'}}\nonumber\\
\label{commutator2}
\end{eqnarray}

\begin{eqnarray}
\frac{1}{2}\bracm{\delta\hat{\mathbf{P}}_{L}^\dag\brac{\mathbf{r},\omega}\delta\hat{\mathbf{P}}_{L}\brac{\mathbf{r'},\omega'}+\delta\hat{\mathbf{P}}_{L}\brac{\mathbf{r'},\omega'}\delta\hat{\mathbf{P}}_{L}^\dag\brac{\mathbf{r},\omega}}
=\nonumber\\-i\frac{\hbar}{\pi}\chi^{aH}(\omega,\Omega_d)\brac{S\brac{\omega}+\frac{1}{2}}\delta\brac{\omega-\omega'}\delta\brac{\mathbf{r}-\mathbf{r'}},\nonumber\\
\label{fdt_Omega}
\end{eqnarray}
where \(\chi^{aH}(\omega,\Omega_d)\) is defined by Eq.~\formref{chi_Omega}. And the first relation Eq.~\formref{commutator2} demonstrates the correctness of calculations, since it gives the right commutator, that can be obtained for the radiation with narrow spectrum just from the requirement of preserving commutation relation for the field operators obeying equations with dissipation. The second relation Eq.~\formref{fdt_Omega} is written in the form analogous to FDT Eq.~\formref{fdt1}, but now the drive-induced modification of dissipation parameter \(\chi^{aH}\) is taken into account, and instead of averaged number of "thermal" photons at frequency \(\omega\) the parameter \(S\) appears, that is defined by the expression: 
\begin{widetext}
\begin{equation}
S=\frac{\rho_{33}\brac{\gamma_{21}\abs{\Omega_d}^2+\gamma_{31}\brac{\gamma_{21}^2+\Delta^2}}+\frac{\abs{\Omega_d}^2n_{23}}{\gamma_{32}}\brac{\abs{\Omega_d}^2+\gamma_{32}\gamma_{31}-\Delta^2}}{n_{13}\brac{\gamma_{21}\abs{\Omega_d}^2+\gamma_{31}\brac{\gamma_{21}^2+\Delta^2}}-\frac{\abs{\Omega_d}^2n_{23}}{\gamma_{32}}\brac{\abs{\Omega_d}^2+\gamma_{32}\gamma_{31}-\Delta^2}}.
\label{S1}
\end{equation}
\end{widetext}
It is evident from Eq.~\formref{S1} that the contribution of \(S\) to the spectral density of fluctuations of polarization is not zero due to the presence of nonzero excitations in the atomic system. It is the noise source related to spontaneous emission processes. It finally defines the spectral density of energy of noise radiation in a medium. It is important that the factor \(S\)
depends explicitly on drive-field intensity. For \(\Omega_d=0\) one gets \(S=\rho_{11}/n_{31}\) that is equal to \(n_T(\omega_{31})\) for the medium in thermal equilibrium with reservoir. To express the quantity \(S\) over the temperature for arbitrary drive intensity it is necessary to calculate the redistribution of atoms over levels induced by the drive wave. 
To this end we should specify the relaxation operators \(\hat{R}_{jj}\). Using a simple form Eq.~\formref{gamma_mn} we express \(\hat{R}_{jj}\) in term of equilibrium (in the absence of drive) population distribution \(r_j^T={{\rho}_{jj}^T}/{{\rho}_{11}^T}=exp\brac{-\hbar\omega_{j1}/T}\) and longitudinal relaxation times, defined for transitions \(\cet{i}-\cet{j}\) as \(T_{ij}=\brac{A_{ij}(n_T(\omega_{ij})+1)}^{-1}\), where \(A_{ij}\) are the Einstein coefficients. Finally we get the following expressions for the stationary populations:
\begin{widetext}
\begin{equation}
\frac{\rho_{jj}}{\rho_{11}}=\frac{r_j^T\brac{\frac{1}{T_{21}T_{31}}+\frac{1}{T_{21}T_{31}}+\frac{1}{T_{31}T_{32}}\frac{r_3^T}{r_2^T}}+
\frac{2\abs{\Omega_d}^2}{\gamma_{32}}\brac{\frac{r_2^T}{T_{21}}+\frac{r_3^T}{T_{31}}}}
{\brac{\frac{1}{T_{21}T_{31}}+\frac{1}{T_{21}T_{31}}+\frac{1}{T_{31}T_{32}}\frac{r_3^T}{r_2^T}}+
\frac{2\abs{\Omega_d}^2}{\gamma_{32}}\brac{\frac{1}{T_{21}}+\frac{1}{T_{31}}}},  j=2,3
\end{equation}
\end{widetext} 
Using the simplest relation between transverse and longitudinal relaxation rates, corresponding to the radiation limit, we can state that the condition \(\gamma_{21}<<\gamma_{31}\), following from Eq.~\formref{EIT_condition}, can be realized  if \(T_{21}\gg T_{31}, T_{32}\); \(r_3^T\ll 1\); \(r_3^T\ll r_{2}^T\), at the same time the level \(\cet{2}\) can be well populated \(r_{2}^T\lesssim 1\), that is possible if \(\omega_{21}\ll\omega_{31}\). Nevertheless the level \(\cet{2}\) is devastated \(\rho_{22} \ll\rho_{22}^T\) due to an action of a sufficiently strong drive wave and fast relaxation from the upper level. Otherwise the EIT regime would easily transform from "transparency" regime to "instability" (so-called Amplification Without Inversion (AWI) \cite{Kochar}) just due to populating of level \(\cet{2}\). It is shown that AWI is achieved for narrow range of dissipative parameters \cite{Kochar_Mandel_Radion}.  We propose simple condition \(T_{31}\approx T_{32}\),  under which the AWI is impossible. Then  
for the strict resonance \(\Delta=0\) and Eq.\formref{EIT_condition} we get the following expression:

\begin{equation}
S=\frac{2r_2^T+\frac{T_{21}}{T_{31}}r_3^T}{1-r_2^T+\frac{T_{21}}{T_{31}}\frac{r_3^T}{r_2^T}}=\frac{2A_{21}n_T(\omega_{21})+A_{31}n_T(\omega_{31})}{A_{21}+A_{31}n_T(\omega_{32})}.
\label{S2}
\end{equation}
Under the same conditions for \(\chi^{aH}_{EIT}\) we get:
\begin{equation}
\chi^{aH}_{EIT}=\frac{i\abs{d_{31}}^2}{\hbar}\frac{1-r_2^T+\frac{T_{21}}{T_{31}}\frac{r_3^T}{r_2^T}}{2T_{21}\abs{\Omega_d}^2}.
\end{equation}
The remarkable point is that due to the process of drive induced parametrical transformation of noise the parameter \(S\) is defined by the number of thermal photons at low frequency \(\omega_{21}\). In the absence of thermal photons at high frequency (\(T\ll\hbar\omega_{31}\)) (but finite spontaneous emission rate at low frequency transition \(A_{21}\neq 0\)) we get a simple relation:
\begin{equation}
S=2n_T\brac{\omega_{21}}.
\label{S3}
\end{equation} 
It is noteworthy that the parameter \(S\), that defines the number of spontaneously emitted photons at high probe frequency in driven medium, can be much larger than unity. 
For physical parameters typical, for example, for Rb vapour widely used in experiments  the formula Eq.~\formref{S3} can be applied for estimation of noise polarization level at probe frequencies (\(\lambda=794 nm\)) even for home temperature. In accordance with this formula the thermal noise, defined by the number of thermal photons at splitting frequency \(\omega_{21}=2\pi \times 6.83 GHz\), becomes significant at temperature \(T\sim 1K\).

It follows from Eq.~\formref{S2} that under ideal condition of zero spontaneous relaxation rate \(A_{21}=0\) the averaged number of noise photons will be defined by the parameter 
\begin{equation}
S=exp(-\hbar\omega_{21}/T),
\label{S4}
\end{equation}
that is less than unity but much higher than the same factor for a number of spontaneously emitted photons in the absence of drive \(exp(-\hbar\omega_{21}/T)>>exp(-\hbar\omega_{31}/T)\). Eq.~\formref{S4} should be used also for the limit of zero splitting frequency \(\omega_{21}\rightarrow 0\), since at that \(A_{21}\rightarrow 0\), \(n_T(\omega_{21})\rightarrow \infty\), but \(A_{21}n_T(\omega_{21})\rightarrow 0\), so that we get \(S(\omega_{21}\rightarrow 0)\rightarrow 1\). But beyond the radiation approximation the presented approach requires an adjustment for being valid at the limit \(\omega_{21}\rightarrow 0\). The problem is that the relaxation model Eq.~\formref{gamma_mn} should be modified if so called secular approximation is violated \cite{JL}, that will induce modification of the correlation functions for the noise operators.    

{\it In summary}, we have revealed the effect of significant field-induced modification of the quantum radiation noise in a thermal reservoir. For the three-level scheme with resonant driving and long-lived low-frequency coherence the obtained relation predicts that in the EIT medium the thermal noise level for the radiation in stationary regime can be significantly higher than in the linear regime without driving, since it is determined by an averaged number of thermal photons at low frequency. But it is necessary to keep in mind that the time (or depth) of establishing this noise solution being determined by the reduced (EIT) decrement is much longer than it is in a linear media.

\begin{acknowledgments}
The authors are grateful to A. Belyanin, I. D. Tokman, V. A. Mironov and V. Vdovin for their useful comments and advices. This work was supported by RFBR grants No. 13-02-00376, No. 13-02-97039, No. 14-22-02034 and No. 14-29-07152.
\end{acknowledgments}

\bibliography{Erukhimova}% Produces the bibliography via BibTeX.

\end{document}